\documentstyle[12pt]{article}
\input epsf.sty
\newcommand{\nH}{n_+}
\newcommand{\nh}{n_-}

\newcommand{\bpt}{\mbox{\boldmath $p$}_{T}}

\newcommand{\bkt}{\mbox{\boldmath $k$}_{T}}
\newcommand{\bqt}{\mbox{\boldmath $q$}_{T}}
\newcommand{\bS}{\mbox{\boldmath $S$}}

\newcommand{\bP}{\mbox{\boldmath $P$}}
\newcommand{\bSt}{\mbox{\boldmath $S$}_{T}}

\newcommand{\xbj}{x_{\scriptscriptstyle B}}

\newcommand{\ba}{\begin{eqnarray}}
\newcommand{\ea}{\end{eqnarray}}
\newcommand{\be}{\begin{equation}}
\newcommand{\ee}{\end{equation}}

\begin{document}

\title{\bf
Opportunities in (semi-)inclusive\\ deep inelastic scattering
}

\vspace{1 cm}
\author{
P.J. Mulders\\
\mbox{}\\
National Institute for Nuclear Physics and High--Energy
Physics (NIKHEF)\\
P.O. Box 41882, NL-1009 DB Amsterdam, the Netherlands\\
and\\
Department of Physics and Astronomy, Free University \\
De Boelelaan 1081, NL-1081 HV Amsterdam, the Netherlands
}
\date{}
\maketitle

\vspace{10 cm}
\noindent
October 1995\\
NIKHEF 95-054\\
nucl-th/9510035

\vspace{3 cm}
\noindent
Talk presented at the ELFE Summer School on Confinement Physics,
22-28 July 1995, Cambridge, England

\newpage
\setcounter{page}{1}

\mbox{}
\vspace{3 cm}
\begin{center}
{\Large{\bf
OPPORTUNITIES IN (SEMI-)INCLUSIVE\\ DEEP INELASTIC SCATTERING}}\\
\vspace{1.5 cm}
{\large P.J. Mulders}\\
\medskip
{\it National Institute for Nuclear Physics and High--Energy Physics (NIKHEF)\\
P.O. Box 41882, NL-1009 DB Amsterdam, the Netherlands}\\
and\\
{\it Department of Physics and Astronomy, Free University \\
De Boelelaan 1081, NL-1081 HV Amsterdam, the Netherlands}
\\
\vspace{3.0 cm}
\large{\bf ABSTRACT}
\end{center}

Inclusive and semi-inclusive deep inelastic leptoproduction offers
possibilities to study details of the quark and gluon structure
of the hadrons involved. In many of these experiments polarization
is an essential ingredient. We also emphasize the dependence on
transverse momenta of the quarks, which leads to azimuthal asymmetries
in the produced hadrons.

\vspace{1.0 cm}

\section{Introduction}

Hard processes using electroweak probes are very well suited to probe the
quark and gluon structure of hadrons. The leptonic part is known,
determining the kinematics of the electroweak probe. Examples of such
processes are
\begin{itemize}
\item
Lepton-hadron scattering (DIS)
\[
\gamma^\ast(q) + H \ \longrightarrow \ h + X \qquad \qquad
(-q^2 \equiv Q^2 \ge 0)
\]
\item
Drell-Yan scattering (DY)
\[
H_A + H_B \ \longrightarrow \ \gamma^\ast(q) + X \qquad \qquad
(q^2 \equiv Q^2 \ge 0)
\]
\item
Electron-positron annihilation
\[
\gamma^\ast(q) \ \longrightarrow \ h_1 + h_2 + X \qquad \qquad
(q^2 \equiv Q^2 \ge 0)
\]
\end{itemize}
The interaction of the electroweak probe with quarks is known.

We consider deep inelastic processes where $Q$ is considerably larger (how
much is mostly an empirical fact) than the typical hadronic scale $\Lambda$,
which is of order 1 GeV. The large momentum $Q$ makes it feasible to
do the calculation within the framework of QCD.
One writes down a diagrammatic expansion of the hard scattering amplitude
containing hard and soft parts. The photon couples into the hard part,
containing quark and gluon lines, while hadrons couple into soft parts,
represented by a blob connecting hadron lines and quark and gluon
lines for which the momenta satisfy $p_i\cdot p_j \sim \Lambda^2 \ll Q^2$. For
the calculation of the hard part one can use the QCD Feynman rules, while for
the soft parts simply the definition enters, being expectation values of quark
and gluon fields in hadron states.

\begin{figure}[hbt]
\begin{center}
\leavevmode
\epsfxsize=14 cm
\epsfbox{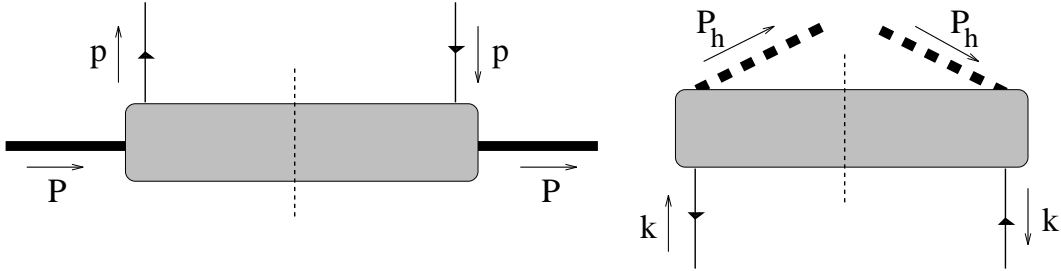}
\end{center}
\caption{\label{fig1}
Quark-quark correlation function giving quark distributions
(left) and fragmentation functions (right)}
\end{figure}
It turns out that at tree level the leading diagrams contain soft parts that
are quark-quark correlation functions of the type shown in Fig.~\ref{fig1},
given by \cite{Soper-77,Collins-Soper-82,Jaffe-83}
\be
\Phi_{ij}(p,P,S) = \frac{1}{(2\pi)^4}\int d^4x\ e^{i\,p\cdot x}
\langle P,S \vert \overline \psi_j(0) \,\psi_i(x)
\vert P,S \rangle,
\ee
where a summation over color indices is implicit, and
\be
\Delta_{ij}(k,P_h,S_h) = \sum_X \frac{1}{(2\pi)^4}\int d^4x\ e^{ik\cdot x}\,
\langle 0 \vert \psi_i(x) \vert P_h, S_h; X \rangle
\langle P_h, S_h;X \vert \overline \psi_j(0) \vert 0 \rangle
\ee
where an averaging over color indices is implicit.
In both definitions flavor indices are suppressed and also the path ordered
link operator needed to make the bilocal matrix element color gauge-invariant
is omitted.

\begin{figure}[hbt]
\begin{center}
\leavevmode
\epsfxsize=14 cm
\epsfbox{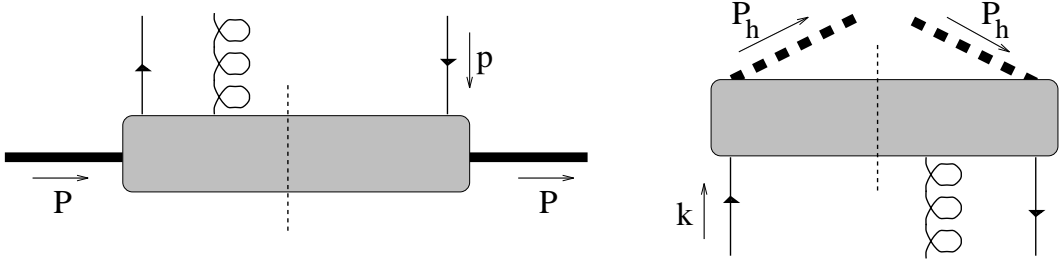}
\end{center}
\caption{\label{fig2}
Quark-quark-gluon correlation functions contributing in hard
scattering processes at subleading order.}
\end{figure}
The large scale $Q$ leads to an ordering of the terms in the diagrammatic
expansion \cite{Ellis-Furmanski-Petronzio-83} in powers of $1/Q$,
$\alpha_s$ and $\alpha_s\ln Q^2$.
Writing down the simplest diagram where a photon is absorbed on a quark
one ends up with the combination of soft parts in Fig.~\ref{fig1}.
Gluonic corrections in the hard QCD part of the process
can be absorbed in a scale dependence of the soft parts, at
least at leading order (factorization). At order $1/Q$ also
quark-quark-gluon correlation functions (shown in Fig.~\ref{fig2}) appear.
These can be rewritten in quark-quark correlation functions using the
QCD equations of motion, provided that one does include the dependence
on the transverse momenta of the quarks.

Next step is the analysis of the correlation functions including the
transverse momentum dependence \cite{Ralston-Soper-79,Tangerman-Mulders-95a}.
It is convenient to parametrize the momenta in terms of lightcone coordinates,
$p = [p^-,p^+,\bpt]$ with $p^\pm = (p^0\pm p^3)/\sqrt{2}$. Choosing a frame in
which the hadrons are collinear one writes for the hadrons and virtual photon
in
$\ell H$ scattering,
\ba
P & \ = \ & \left[ \frac{\xbj M^2}{A\sqrt{2}},
\frac{A}{\xbj \sqrt{2}}, \mbox{$\bf 0$}_{T} \right]
 \ \equiv \  \frac{Q}{\xbj\sqrt 2}\,\nH + \frac{\xbj M^2}{Q\sqrt{2}}\,\nh,
\label{collinearP} \\
P_h & \ = \ & \left[ \frac{z_h Q^2}{A\sqrt{2}},
\frac{A M_h^2}{z_h Q^2 \sqrt{2}},
\mbox{$\bf 0$}_{T} \right]
\ \equiv \  \frac{z_h\,Q}{\sqrt 2}\,\nh + \frac{M_h^2}{z_h\,Q\sqrt{2}}\,\nH,
\label{collinearPh} \\
q & \ = \ & \left[ \frac{Q^2}{A\sqrt{2}}, -\frac{A}{\sqrt{2}},
\bqt \right]
\ = \ \frac{Q}{\sqrt 2}\,\nH - \frac{Q}{\sqrt{2}}\,\nh
+ q_{\scriptstyle T}.
\label{collinearq}
\ea
Note that in a frame in which $P$ and $q$ have no transverse momenta,
the outgoing hadron has a transverse momentum $\bP_{h\perp} = -z\bqt$.
The calculation of the diagrams involves an integral over soft parts,
\ba
\Phi^{[\Gamma]}(x,\bpt) & = &
\left. \frac{1}{2}\int dp^-\ Tr(\Phi\,\Gamma) \right|_{p^+ =
x P^+,\ \bpt}, \\
\Delta^{[\Gamma]}(z,\bkt) & = &
\left. \frac{1}{4z}\int dk^+\ Tr(\Delta\,\Gamma)\right|_{k^- =
P_h^-/z,\ \bkt} .
\ea
Depending on the Dirac matrix $\Gamma$, these correlation functions
are parametrized in terms of distribution and fragmentation
functions, e.g. for a polarized spin 1/2 target with spin vector
$S = \left[-\lambda\,M/2P^+, \lambda\,P^+/M,\bS_T\right]$ with $\lambda^2 +
\bS_T^2$ = 1,
\begin{eqnarray}
& & \Phi^{[\gamma^+]}(x,\bpt) = f_1(x ,\bpt) ,
\\ & & \Phi^{[\gamma^+ \gamma_5]}(x,\bpt) =\lambda\,g_{1L}(x ,\bpt)
+ g_{1T}(x ,\bpt)\,\frac{(\bpt\cdot\bSt)}{M} \equiv g_{1s}(x,\bpt) ,
\\ & & \Phi^{[ i \sigma^{i+} \gamma_5]}(x,\bpt) =
S_{\scriptstyle T}^i\,h_{1T}(x ,\bpt)
+ \frac{p_{\scriptstyle T}^i}{M}\,h_{1s}^\perp (x,\bpt), \\
& & \Phi^{[1]}(x,\bpt) = \frac{M}{P^+}\,e(x ,\bpt)
\\ & & \Phi^{[\gamma^i]}(x,\bpt) =
\frac{p_{\scriptstyle T}^i}{P^+}\,f^\perp(x ,\bpt),
\\ & & \Phi^{[ \gamma^i \gamma_5]}(x,\bpt) =
\frac{M\,S_{\scriptstyle T}^i}{P^+}
\, g_{\scriptstyle T}^\prime(x ,\bpt)
+ \frac{p_{\scriptstyle T}^i}{P^+}\,g_{s}^\perp(x,\bpt)
\\ & & \Phi^{[i \sigma^{ij} \gamma_5]}(x,\bpt) =
\frac{S_{\scriptstyle T}^ip_{\scriptstyle T}^j
-p_{\scriptstyle T}^iS_{\scriptstyle T}^j}{P^+}
\,h_{\scriptstyle T}^\perp(x ,\bpt)
\\ & & \Phi^{[ i\sigma^{+-}\gamma_5 ]}(x,\bpt) =
\frac{M}{P^+} \,h_{s}(x ,\bpt).
\end{eqnarray}
In naming the functions we have extended the scheme proposed by Jaffe and
Ji \cite{Jaffe-Ji-92} for the $\bkt$-integrated functions.
Depending on the Lorentz structure of the Dirac matrices $\Gamma$ the
parametrization involves powers $(1/P^+)^{t-2}$, where $t$ is referred to as
'twist'. Integrated over $k_T$ and taking moments in $x$ it corresponds to the
OPE 'twist' of the (in that case) local operators. When everything is done it
will turn out that the factors $1/P^+$ give rise to factors $1/Q$ in the cross
sections. The leading projections $\Phi^{[\gamma^+]}$,
$\Phi^{[\gamma^+\gamma_5]}$ and $\Phi^{[i\sigma^{+i}\gamma_5]}$ can be
interpreted as quark momentum densities, namely the unpolarized distribution,
the chirality (for massless quarks helicity) distribution and the transverse
spin distribution, respectively.

For the fragmentation functions one has an analogous analysis, which for
unpolarized final state hadrons yields
\begin{eqnarray}
& & \Delta^{[\gamma^-]}(z,\bkt) = D_1(z,-z\bkt) ,
\\ & & \Delta^{[i \sigma^{i-} \gamma_5]}(z,\bkt) =
\frac{\epsilon_{\scriptscriptstyle T}^{ij} k_{T j}}{M_h}\,H_1^\perp(z,-z\bkt),
\\ & & \Delta^{[1]}(z,\bkt) =
\frac{M_h}{P_h^-}\,E(z,-z\bkt) ,
\\ & & \Delta^{[\gamma^i]}(z,\bkt) =
\frac{k_T^i}{P_h^-}\,D^\perp(z,-z\bkt),
\\ & & \Delta^{[ i \sigma^{ij} \gamma_5]}(z,\bkt) =
\frac{M_h\,\epsilon_{\scriptscriptstyle T}^{ij}}{P_h^-}\,H(z,-z\bkt).
\end{eqnarray}
Here each power $1/P_h^-$ leads to a factor $1/Q$ in the cross section. The
functions $H_1^\perp$ and $H$ have no equivalent for distribution functions.
They are allowed for the fragmentation functions because time reversal
invariance cannot be used in the analysis for $\Delta$ which involves
out-states $\vert P_h, X\rangle$.

Putting everything together \cite{Mulders-Tangerman-96},
the result of the tree-level calculation up
to order $1/Q$ involves combinations of the distribution and fragmentation
functions defined above. The inclusion of qqG-correlation functions
of the type in Fig.~\ref{fig2} and their relation to qq-correlations through
the equations of motion are essential to ensure electromagnetic gauge
invariance.
We give 2 specific examples of cross sections. The first is well-known, being
the result for inclusive $\ell H$ scattering up to order $1/Q$ including
polarization. Using the scaling variables $x = Q^2/2P\cdot q$ and
$y = P\cdot k/P\cdot q$ one obtains
\ba
\frac{d\sigma} {d\xbj\,dy} & = &
\frac{4\pi \alpha^2\,s}{Q^4}\ \Biggl\{\,
         \left\lgroup \frac{y^2}{2}+1-y\right\rgroup \xbj f_1(\xbj)
         + \lambda_e\,\lambda
          \,y \left\lgroup 1-\frac{y}{2} \right\rgroup \xbj g_1(\xbj)
\nonumber
\\&& - \lambda_e\,\vert \bS_\perp\vert\,\frac{M}{Q}
            \,2\,y\sqrt{1-y}\, \cos (\phi_s)\,\xbj^2\, g_T(\xbj)\,\Biggr\} .
\ea
As indicated before, the twist-3 function in $\Phi^{[\gamma^i\gamma_5]}$
surviving after $k_T$-integration, $g_T = g_T^\prime +
(\bkt^2/2M^2)\,g_T^\perp$, appears at subleading order. Reinstating
the summation over quark flavors and identifying the result with the most
general cross sections, expressed in terms of structure functions, one obtains
\ba
 &&\frac{F_2(\xbj,Q^2)}{\xbj} = 2\,F_1(\xbj,Q^2) =
\sum_a e_a^2\,\left( f_1^{(a)}(\xbj) + \bar f_1^{(a)}(\xbj) \right),\\
&&2\,\mbox{\large {\em g}}_1(\xbj,Q^2) =
\sum_a e_a^2\,\left( g_1^{(a)}(\xbj) + \bar g_1^{(a)}(\xbj) \right),\\
&&\mbox{\large {\em g}}_T(\xbj,Q^2) =
\mbox{\large {\em g}}_1(\xbj,Q^2) + \mbox{\large {\em g}}_2(\xbj,Q^2) =
\frac{1}{2} \sum_a e_a^2\,\left( g_T^{(a)}(\xbj) + \bar g_T^{(a)}(\xbj)
\right).
\ea

The second example is semi-inclusive scattering including the dependence on the
transverse momentum $\bP_{h\perp}$ of the detected
hadron \cite{Kotzinian-95,Tangerman-Mulders-95b}. For this we assume
a gaussian transverse momentum dependence for the quark distribution
and fragmentation functions,
\ba
f(x,\bpt^2) & = & f(x)\,\frac{R_H^2}{\pi}\,\exp (-R_H^2 \bpt^2)
\equiv f(x)\,{\cal G}(\vert\bpt\vert;R_H), \\
D(z,z^2\bkt^2) & = & D(z)\,\frac{R_h^2}{\pi\,z^2}\,\exp (-R_h^2 \bkt^2)
= \frac{D(z)}{z^2}\,{\cal G}(\vert\bkt\vert;R_h)
= D(z)\,{\cal G}\left(z\vert\bkt\vert ; \frac{R_h}{z}\right).
\ea
The result for the cross section is
\begin{eqnarray}
&&\frac{d\sigma}{d\xbj\,dy\,dz_h\,d^2\bP_{h\perp}} \ = \
\frac{4\pi \alpha^2\,s}{Q^4}\,\sum_{a,\bar a} e_a^2\,
\left\lgroup \frac{y^2}{2}+1-y\right\rgroup  \,\xbj f_1^a(\xbj)\,D^a_1(z_h)
\,\frac{{\cal G}(Q_T;R)}{z_h^2}
\nonumber \\ && \qquad \quad \mbox{}
-\frac{4\pi \alpha^2\,s}{Q^4}\,\lambda\,\sum_{a,\bar a} e_a^2\,
(1-y)\, \sin (2\phi_h)\,\frac{Q_T^2\,R^4}{M M_h\,R_H^2\,R_h^2}
\,\xbj h_{1L}^{\perp\,a}(\xbj) H_1^{\perp\,a}(z_h)
\,\frac{{\cal G}(Q_T;R)}{z_h^2}
\nonumber \\ && \qquad \quad \mbox{}
-\frac{4\pi \alpha^2\,s}{Q^4}\,\vert \bS_\perp\vert
\,\sum_{a,\bar a} e_a^2\,\Biggl\{
(1-y)\,\sin(\phi_h + \phi_s) \,\frac{Q_T\,R^2}{M_h\,R_h^2}
\,\xbj h^a_1(\xbj) H_1^{\perp\,a}(z_h)
\nonumber \\ && \qquad\qquad\qquad\quad
+(1-y) \,\sin(3\phi_h - \phi_s) \,
\frac{Q_T^3\,R^6}{2M^2M_h\,R_H^4\,R_h^2}
\,\xbj h_{1T}^{\perp\,a}(\xbj) H_1^{\perp\,a}(z_h)
\Biggr\}\,\frac{{\cal G}(Q_T;R)}{z_h^2}
\nonumber \\ && \qquad \quad \mbox{}
+\frac{4\pi \alpha^2\,s}{Q^4}\,\lambda_e\lambda\,\sum_{a,\bar a} e_a^2\,
 y\left(1-\frac{y}{2}\right) \,\xbj\,g^a_{1L}(\xbj)\,D^a_1(z_h)
\,\frac{{\cal G}(Q_T;R)}{z_h^2}
\nonumber \\ && \qquad \quad \mbox{}
+\frac{4\pi \alpha^2\,s}{Q^4}\,\vert \bS_\perp\vert
\,\sum_{a,\bar a} e_a^2\,
y\left(1-\frac{y}{2}\right)\,\cos (\phi_h - \phi_s)\,\frac{Q_T\,R^2}{M\,R_H^2}
\,g^a_{1T}(\xbj) D^a_1(z_h)
\,\frac{{\cal G}(Q_T;R)}{z_h^2}.
\end{eqnarray}
We see that all six twist-two $x$- and $\bpt$-dependent quark distribution
functions for
a spin 1/2 hadron can be accessed in leading order asymmetries if one
considers lepton and hadron polarizations. One of the asymmetries involves
the transverse spin distribution $h_1^a$ \cite{Collins-93}.
On the production side, only two
different fragmentation functions are involved, the familiar unpolarized
fragmentation function $D_1^a$ and the (interaction-dependent) fragmentation
function $H_1^{\perp a}$.

This work is part of the research program of the foundation for Fundamental
Research of Matter (FOM) and the National Organization for Scientific
Research (NWO).

\end{document}